\documentclass{article}
\usepackage{spconf,amsmath,graphicx}
\usepackage[utf8]{inputenc}
\usepackage{float}
\usepackage{cite}
\usepackage{amssymb}
\usepackage[vlined,boxed]{algorithm2e}
\usepackage{xcolor}
\usepackage{caption}
\usepackage{subcaption}
\newcommand{\RR}{\ensuremath{\mathbb R}}
\newcommand{\minimize}[2]{\ensuremath{\underset{\substack{{#1}}}{\mathrm{minimize}}\;\;#2 }}
\usepackage[nomargin,inline,draft]{fixme}
\fxsetup{theme=color,mode=multiuser}
\FXRegisterAuthor{isa}{aisa}{\color{cyan}isabela}
\FXRegisterAuthor{me}{ame}{\color{blue}mireille}
\FXRegisterAuthor{pf}{apf}{\color{red}pascal}
\graphicspath{ {images/} }

\title{Distributed Graph Learning with Smooth Data Priors}
\name{Isabela Cunha Maia Nobre\textsuperscript{1}, Mireille El Gheche\textsuperscript{2}, Pascal Frossard\textsuperscript{1}\thanks{Isabela Nobre is partially supported by CAPES (grant number 88881.174577/2018-01).}}
\address{\textsuperscript{1}EPFL, LTS4, Lausanne, Switzerland\\ \textsuperscript{2}Sony AI, Z\"urich, Switzerland}
\begin{document}
%
\maketitle
\begin{abstract}
Graph learning is often a necessary step in processing or representing structured data, when the underlying graph is not given explicitly. Graph learning is generally performed centrally with a full knowledge of the graph signals, namely the data that lives on the graph nodes. However, there are settings where data cannot be collected easily or only with a non-negligible communication cost. In such cases, distributed processing appears as a natural solution, where the data stays mostly local and all processing is performed among neighbours nodes on the communication graph. We propose here a novel distributed graph learning algorithm, which permits to infer a graph from signal observations on the nodes under the assumption that the data is smooth on the target graph. We solve a distributed optimization problem with local projection constraints to infer a valid graph while limiting the communication costs. Our results show that the distributed approach has a lower communication cost than a centralised algorithm without compromising the accuracy in the inferred graph. It also scales better in communication costs with the increase of the network size, especially for sparse networks. 
\end{abstract}
\section{Introduction}

Structured data, which is defined over the nodes of a graph or network, is very common nowadays in machine learning, data science and signal processing fields. Much of the data in computer or transportation systems, energy networks, social, biological and other scientific applications can be represented with a graph. This situation has fostered the development of effective operators to process such irregularly structured data, which are often grouped under the umbrella of graph signal processing (GSP). However, there are settings where the underlying graph is not given explicitly and has to be learned, which has led to the so-called problem of graph learning (GL)\cite{natali2020topology,minguez2020state,roddenberry2020network,thanou2017learning,lassance2020graph,matta2019graph,minder2020figlearn,ying2020does}. Most graph learning methods assume that data can be gathered at no cost and that the problem can be solved centrally. This assumption is however challenged in settings where data actually lives on physically separated nodes and where collecting this information induces communication costs. Unfortunately, these costs are generally overlooked in the graph learning literature. 

In this work, we propose the first distributed graph learning framework that carefully considers the communication costs for exchanging data between the different nodes of the graph. We assume that there is a communication graph that connects the nodes, and we search for a data graph that appropriately describes the structure of the data. We formulate a graph learning problem under the assumption that the data varies smoothly over the target data graph. Then, we propose a new distributed optimisation algorithm to solve the graph learning problem while limiting the communication costs. This is achieved by local processing in the nodes, along with distributed projections to satisfy the optimization constraints. Our experiments show that the communication costs of our distributed algorithm is lowered without compromising the accuracy of the inferred graph, compared to a centralized graph learning algorithm. Besides, it scales better in communication costs with an increase of the network size or of the number of graph signals. We also show that sparse networks benefit more from a distributed solution than dense ones. Our new algorithm certainly opens the way to other distributed graph learning solutions for dynamic settings where graph signals evolve with time. 

Some other works have considered the graph learning problem in the context of distributed algorithms. For instance, the authors in \cite{matta2020graph,matta2018consistent} consider networks that perform decentralized processing tasks. They propose to observe the evolution of the signals in order to infer the underlying graph that lead to the observed signal evolution. The graph learning task in itself, however, requires a central processor that gathers all observed signals. Similarly, the authors in \cite{de2013joint} present a fusion center with the objective of joint recovery of network transmitted data and connection topology in a collaborative network. Closer to our work, the works \cite{moscu2019learning,moscu2020online} aim to learn a graph with a graph-based filtering multivariate model. None of the above works however takes into account the communication costs in the process of learning the graph. This is exactly the motivation for our work.

\section{Problem formulation}

\subsection{Graph Learning}

We consider a weighted undirected graph, which can be written as $\mathcal{G}_d=(\mathcal{V},\mathcal{E}, W)$, where $\mathcal{V}$ represents a set of $N$ vertices, $ \mathcal{E} $ represents a set of edges, and $W \in \mathbb{R}^{N\times N}$ is the weight matrix whose entries $w_{i,j}$ represent the pairwise relationship between nodes. Given the degree matrix $D$, whose diagonal entries equal $D_{ii} = d_i = \sum_{j}w_{i,j}$, and zero outside the diagonal, we can also define the Laplacian matrix as  ${L} = D-W  $. A graph signal is a function defined on the vertices of this graph. Assuming each node $i$ contains $M$ features, it can be associated with a feature vector $x_i \in  \mathbb{R}^{M}$. This is equivalent to say that the whole graph has $M$ graph signals, which can then be represented as $X \in \mathbb{R}^{N\times M}$. 

Equipped with the smoothness assumption\footnote{The smoothness can be computed with the Laplacian quadratic form 
\begin{equation*}
\mathcal{Q}({L}) = \text{tr} 	(X^T{L}  X) =  \frac{1}{2} \sum_{i,j}w_{i,j}||x_i-x_j||^2.
\end{equation*}}, one can formulate the graph learning problem as follows:
\begin{equation}
\label{eq:prob_formulation}
\minimize{{L} \in \mathcal{C}}{\text{tr}(X^T {L} X)},
\end{equation}
where $\mathcal{C}$ is the space of valid combinatorial graph Laplacian matrices defined as
\begin{equation}
\mathcal{C} = \{L\in\mathbb{R}^{N\times N} \,|\, \; L_{i,j}=L_{j,i}\leq 0, L_{i,i}=-\sum_{j\neq i} L_{i,j}\}.
\end{equation}
To facilitate the optimization defined in Equation \eqref{eq:prob_formulation} with respect to the graph structure, we seek for a valid weight matrix\footnote{$\mathcal{W} = \{W\in\mathbb{R}^{N\times N} \,|\, \; W=W^\top, {\rm diag}(W)=0, w_{i,j} \geq 0\}.$} instead of a Laplacian. Moreover, we can exploit the symmetry of the adjacency matrix by only optimizing the upper triangular part $w \in \mathcal{W}$. If we denote as $\mathcal{L}\colon \RR^{(N-1)N/2} \rightarrow \RR^{N\times N}$ the linear operator that converts the upper-triangular part $w$ of a weight matrix into the corresponding Laplacian matrix, we can reformulate the problem as
\begin{equation}
  \minimize{w \in [0,1]^{(N-1)N/2}}{ \text{tr}  (X^\top \mathcal{L}w X)}.
\end{equation}




\subsection{Distributed setup and problem formulation}


We now consider a network, for instance a wireless sensor network, whose nodes have a communication range that defines which other nodes they can directly exchange messages with. It defines an unweighted communication graph $\mathcal{G}_c$ over the vertex set $ \mathcal{V} $. We also assume that the unknown data graph $\mathcal{G}_d$ is a spanning subgraph of the communication graph.  The nodes do not have direct access to a central unit: in order to learn the data graph $\mathcal{G}_d$, they have to use $\mathcal{G}_c$  for communication and processing, under the restriction that nodes can only communicate directly with 1-hop neighbors on $\mathcal{G}_c$. Finally, each message transmitted on the communication graph eventually induces communication costs.

We propose to learn the data graph in a distributed way, in order to lower the communications costs compared to a centralised solution. It means that each node $n$ should learn which neighbor they have on the data graph $\mathcal{G}_d$ and the corresponding weight of these connections. 
Since the data graph is a spanning subgraph of the communication graph, the potential connecting neighbours $j$, for node $i$ in the data graph, should be picked up from $\mathcal{G}_c$. Equivalently, it has to belong to the neighborhood of $i$, represented by $\mathcal{N}_i^c$. We can then rewrite the graph learning problem presented above, but in distributed settings. For node $i$, the distributed graph learning problem thus reads:
\begin{equation}\label{problem_formulation2}
 \minimize{w \in [0,1]^{(N-1)N/2}}{\sum_i \sum_{j\in \mathcal{N}_i^c} (X^\top \mathcal{L}w X)  
}
\end{equation}
In a distributed setting, solving Equation \eqref{problem_formulation2} is not trivial while minimizing the communication cost between the nodes. Hence, we reformulate the problem as:
\begin{align}
\minimize{w_{i,j} \geq 0}{ & \sum_{i,j\in\mathcal{N}_j^c} w_{i,j} \|x_i-x_j\|^2} \label{eq:objective_local}\\ 
& + \sum_i {\rm max} \{0, \eta - \sum_{j\in\mathcal{N}_j^c} w_{i,j} \}^2 \label{eq:local} \\  & {\rm s.t.} \quad \, w_{i,j} = w_{j,i}. \label{eq:constraint}
\end{align}
Where $\eta>0$ is the desired minimum degree for every node. To solve the graph learning problem in a distributing setting, we 
propose to learn $\mathcal{G}_d$ while limiting the communication costs. The algorithm consists of a collaborative learning between the graph nodes: local minimization to compute the weights $w_{i,j}$, $\forall i\in \{1,\dots,N\}$
\begin{equation}\label{problem_formulationdist}
    \minimize{w_{i,j} \geq 0}{\frac{1}{N} \sum_{j\in\mathcal{N}_i^c} w_{i,j} \|x_i-x_j\|^2 + {\rm max} \{0, \eta - \sum_{j\in \mathcal{N}^c_i} w_{i,j} \}} ^2
\end{equation}
and global optimization to maintain the symmetry constraint as detailed in (\ref{eq:constraint}). 

\section{Distributed Graph Learning Algorithm}

We detail now the algorithm for solving the graph learning problem in a distributed way. 
The problem can be locally solved by two steps: an initialization, where each node $i$ communicates its signal $x_i$ to the neighbors $j\in \mathcal{N}_i^c$; and an optimization where the nodes collaboratively find the solution of \eqref{eq:objective_local}, \eqref{eq:local}, and \eqref{eq:constraint} by exchanging the local estimates of the graph weight values. We describe these two steps in more details below.

\paragraph*{Initialization}

\begin{figure}[htb]
     \centering
     \begin{subfigure}[b]{.48\linewidth}
         \centering
         \centerline{\includegraphics[width=4.0cm]{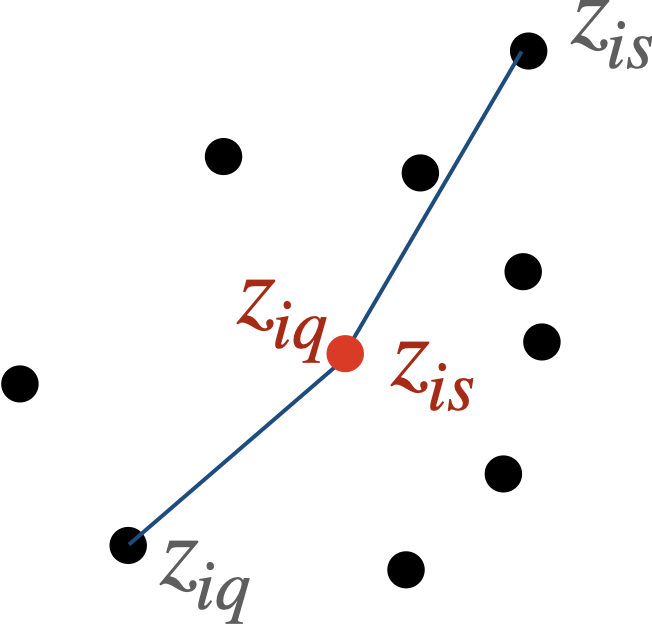}}
         \caption{}
         \label{fig:toynetwork3}
     \end{subfigure}
          \hfill
     \begin{subfigure}[b]{.48\linewidth}
         \centering
         \centerline{\includegraphics[width=4.0cm]{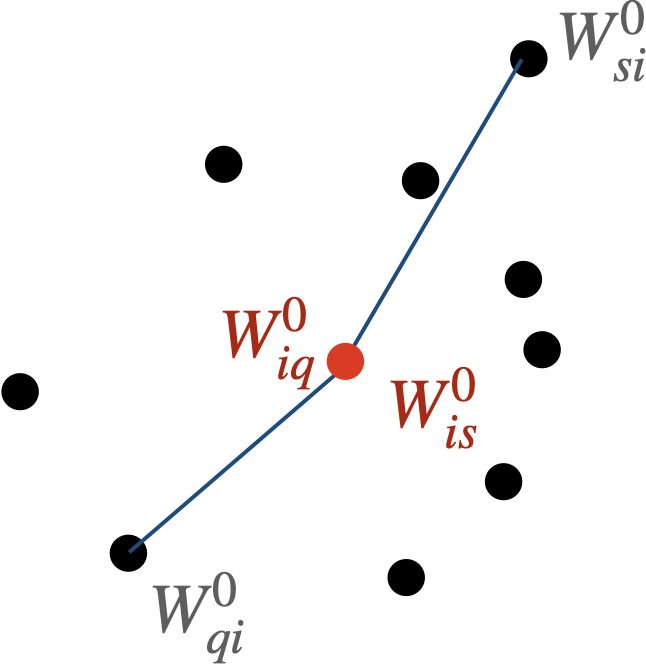}}
         \caption{}
         \label{fig:toynetwork4}
     \end{subfigure}
          \hfill
     \begin{subfigure}[b]{.48\linewidth}
         \centering
         \centerline{\includegraphics[width=4.0cm]{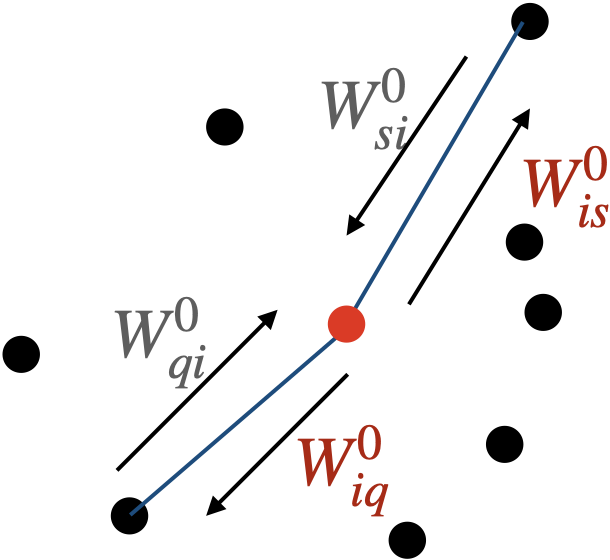}}
         \caption{}
         \label{fig:toynetwork5}
     \end{subfigure}
          \hfill
     \begin{subfigure}[b]{.48\linewidth}
         \centering
         \centerline{\includegraphics[width=4.0cm]{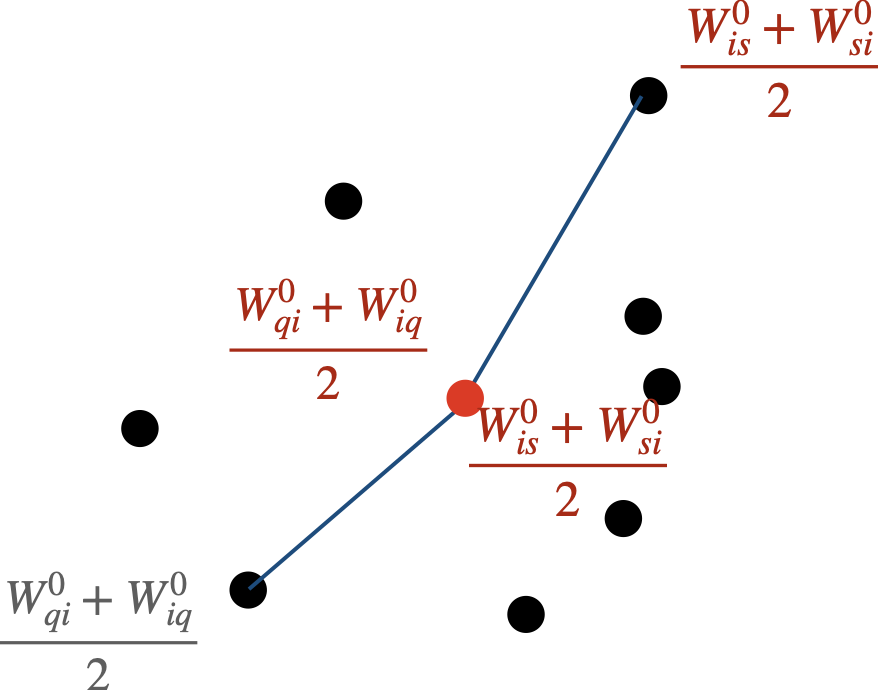}}
         \caption{}
         \label{fig:toynetwork6}
     \end{subfigure}
          \hfill
     \begin{subfigure}[b]{.48\linewidth}
         \centering
         \centerline{\includegraphics[width=4.0cm]{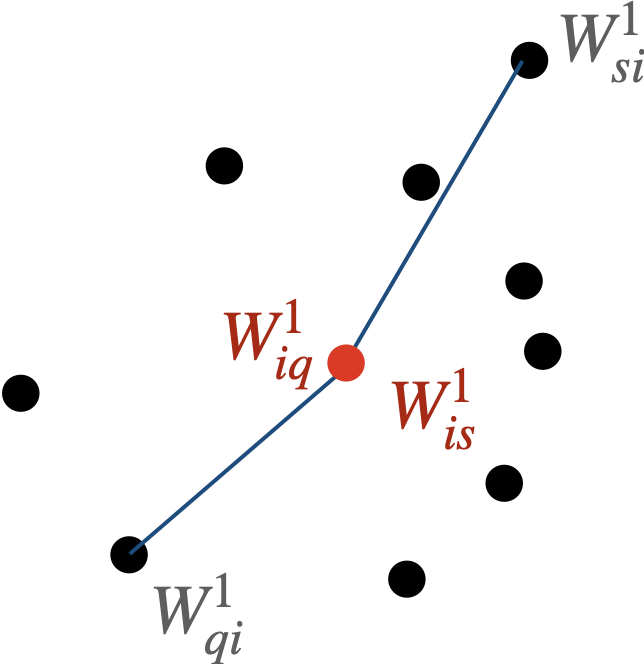}}
         \caption{}
         \label{fig:toynetwork7}
     \end{subfigure}
        \caption{Toy network illustrating the algorithm.  \ref{fig:toynetwork3}: After initialization node $i$ obtains the difference vectors. \ref{fig:toynetwork4}: First estimation of the data graph weights. \ref{fig:toynetwork5}: Sharing data graph weights. \ref{fig:toynetwork6}: Weights averaging, or symmetry constraint projection. \ref{fig:toynetwork7}: One step updating of gradient descent.} 
        \label{toynetworksall2}
\end{figure}

We propose an initialization algorithm for obtaining the values of $z_{ij}=\|x_i - x_j\|^2$ at each node $i$ (Fig. \ref{fig:toynetwork3}). The algorithm takes advantage of the properties of the communication graph in trying to limit communication costs. It is constructed as follows. While there are still $z_{ij}$ to be calculated, every node $i$ sends $x_i$ to its neighbour with the highest degree, out of those whose corresponding $z_{ij}$ has not yet been determined, as long as it is larger or equal to the degree of node $i$ and $i$ did not receive the signals from that node. If these conditions are not fulfilled, $x_i$ is not sent. The nodes that have received the $x_i$ values then calculate all possible combinations of $  z_{ij}  $ for themselves and for their neighbours. The corresponding results are sent to these neighbours. A similar round happens again, for the nodes that still do not have the $z_{ij}$ of all its neighbours, and repeats until all $z_{ij}$ combinations, of edges that belong to $\mathcal{G}_c$, are calculated. The algorithm lowers the communication costs by reducing the amount of times the $x_i$ values are sent.

\paragraph*{Optimization}
The positivity in (\ref{problem_formulationdist}) is local and describe closed sets, so each node can independently project the solution into the set defined by these constraint using ReLU function.

The symmetry constraint also represents a closed set, so it can be solved by projection. It cannot be independently solved by each node, requiring communication though. Each node shares its $w_{i,j}$ values with the corresponding neighbours and updates them with $w_{i,j}^{projected} =  \frac{w_{i,j}+w_{j,i}}{2}$. From an optimization point of view, the projection should be done at every step of the algorithm. But since this projection costs in communication, we choose a compromise solution where the nodes first solve their local optimization problem (\ref{problem_formulationdist}), without the symmetry constraint; and, once the local problem converges at every node (figs. \ref{fig:toynetwork4} and \ref{fig:toynetwork7}), they communicate and jointly apply the symmetry constraint (figs. \ref{fig:toynetwork5} and \ref{fig:toynetwork6}). Symmetry projection and optimization alternate until global convergence.

The degree constraint represents an open set, so we replace it by the constraint $\sum_{j}w_{i,j}\ge \eta$, with $ \eta>0$. We propose to add the regularization term $\max\left\lbrace 0,\eta-\sum_{j}w_{i,j}\right\rbrace ^{2}$  to the objective function. The term is differentiable, which enables a solution with gradient descent. It prevents the degree to become too small by penalizing the solution when $d$ becomes smaller than $\eta$. It does not affect the solution when $d$ is equal or bigger than $\eta$, since the regularization term becomes zero in that case. One advantage compared to other solutions (e.g., logarithm function) is that when $d$ approaches zero, the regularization term does not go to infinity, bringing stability to the solution. The cost function we solve at node $i$ becomes then (\ref{problem_formulationdist}), with the data fidelity term normalized by the size of the network. 

One of the advantages of our method is that is has a very simple objective function, without any parameters to be set. This is appropriate for a distributed scenario, since the task of distributively setting up parameters is a problem on its own, which would also involve extra communication costs. Note that the variable $\eta$ is not a free parameter, but rather an input of the problem that can be directly adapted to the desired graph sparsity level.

In Algorithm \ref{alge} we can see the global solution for the distributed graph learning algorithm.

\begin{figure}[htb]
		\centering
\includegraphics[width=\linewidth]{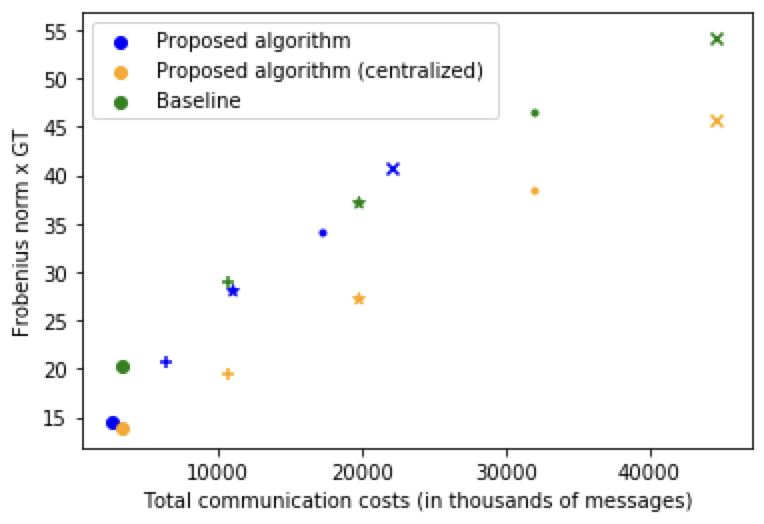} 
\caption{Frobenius norm (of the difference between learned and ground truth weight matrices) versus total communication costs for the three methods in sparse settings. Big circle: $N =150$, + symbol: $N=350$, star: $N=550$, small circle: $N=750$, X symbol: $N=950$.}
\label{fig:FrobxCommSparse}
\end{figure}

\begin{algorithm}[htb]
    Initialization algorithm (Fig. \ref{fig:toynetwork3})\\
    Initialize all weights $w_{i,j} = 1$\\
	\While{the global algorithm does not converge}{
	\For{every node $i$}{
	\While{The local algorithm in node $i$ does not converge}{
	    Relu\\
	    set $w_{i,i} = 0$\\
        Gradient step: minimize (\ref{problem_formulationdist}) 
        }
        }
        After local convergence, local data graph weights are estimated (Figs. \ref{fig:toynetwork4} and \ref{fig:toynetwork7})\\
	\For{every node $i$}{
		Node $i$ gives $w_{i,j}^t$ to every node $j$ that is a neighbor in the communication graph (Fig. \ref{fig:toynetwork5})\; 
		Node $i$ receives $w_{j,i}^t$ from every node $j$ that is a neighbor in the communication graph (Fig. \ref{fig:toynetwork5})\; 
		Node $i$ makes: $ w_{i,j}^t \gets \frac{w_{i,j}^t+w_{j,i}^t}{2}$ for all $j$ (Fig. \ref{fig:toynetwork6})\; 
		
		}
		\For{every node $i$}{
        Relu\\
        set $w_{i,i} = 0$
		}
	}
	\caption{Distributed algorithm global solution}\label{alge}
\end{algorithm}

\section{Experimental Results}
\subsection{Performance}\label{sec:performance}

We consider synthetic settings and create the communication graph by randomly distributing $N$ nodes uniformly inside the unit square and assigning a communication radius value equal to every node. We repeat this process until producing a connected graph. A variable called removal-rate defines the ratio of edges in the communication graph that will be randomly deleted, for the data graph generation, with random weights assigned to the edges. We assign 5000 smooth signals from a probabilistic generative model to each node, the removal-rate we fix at 0.5, and we vary the number of nodes $N$ from 150 to 950 \footnote{Sampled every 200}. In order to account for change in node density, due to the fixed unitary area, we use different communication radius for different $N$ values, with the proportion $\frac{2}{\sqrt{N}}$.\footnote{The square root is used to account for the 2-dimensional nature of the problem.}

We compare our distributed algorithm with the centralized graph learning approach in \cite{kalofolias2017large}, as well as with a centralized version of our algorithm, where (\ref{eq:local}) is solved directly, with the symmetry projection done at every step of the optimization algorithm. The communication cost is the number of messages exchanged between two nodes, where one scalar number equals one message. For our distributed framework, the total cost is the sum of the initialization and the optimization costs.  For the centralized methods, it is the cost to aggregate all signals into a node with lowest eccentricity plus the costs of bringing the results back. The accuracy metrics are defined over the difference between the ground truth graph and the learned ones. We use the Pytorch optimizer Adam, with an independent optimizer for each node in the distributed case, and only one optimizer for the centralized cases. 

\begin{figure}[htb]
	\centering
	\includegraphics[width=\linewidth]{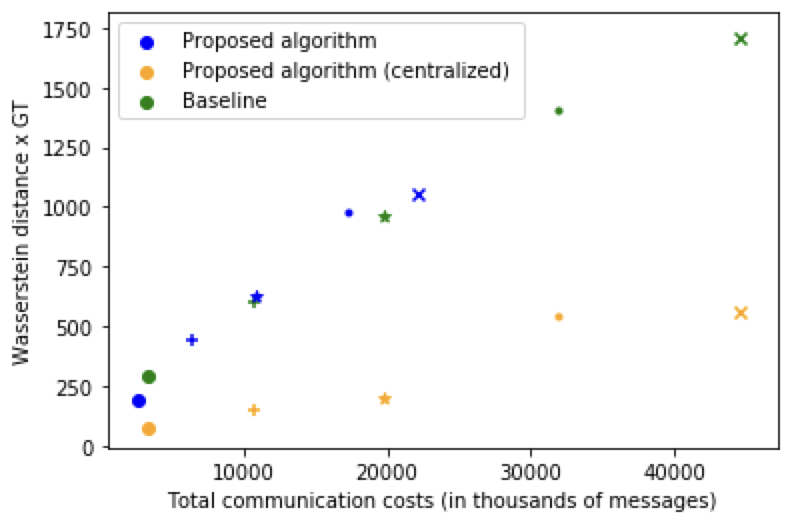}
	\caption{Wasserstein distance (between learned and ground truth weight matrices) versus total communication costs for the three methods in sparse settings. Big circle: $N =150$, + symbol: $N=350$, star: $N=550$, small circle: $N=750$, X symbol: $N=950$.}
	\label{fig:WarsxCommSparse}
\end{figure}

 In Figs. \ref{fig:FrobxCommSparse} and \ref{fig:WarsxCommSparse}  we plot performance in terms of both total communication costs and accuracy (Frobenius and Wasserstein) for the three algorithms. 
 These experiments have been run in the sparse settings (average of 11.57 neighbours/node in communication graph). We notice that our distributed algorithm scales better in communication costs with the increase of $N$. The distributed version presents better communication costs for all cases. In terms of accuracy, the distributed algorithm is only sometimes slightly worse than the centralized counterparts, due to the approximations inherent to its design. The distributed algorithm then outperforms the centralised ones and represents a good trade-off between communication costs and graph estimation accuracy.


In the same configuration as before, we assign 1000 signals to each node. We vary $N$ from 50 to 150. We also assign 4 different values, 0.2, 0.3, 0.4 and 0.5, for the communication radius, and average the results over these values. We compare the communication costs of calculating the differences $z_{ij}$ with both our initialization algorithm and the naive approach where each node send each other their respective signals in order to calculate the difference values.  We note that our initialization algorithm not only presents lower communication costs, it also scales well with the size of the network.


\subsection{Analysis}

Here we study in details the performance of the algorithm with respect to different parameters. 

\begin{figure}[htb]
		\centering
	\includegraphics[width=\linewidth]{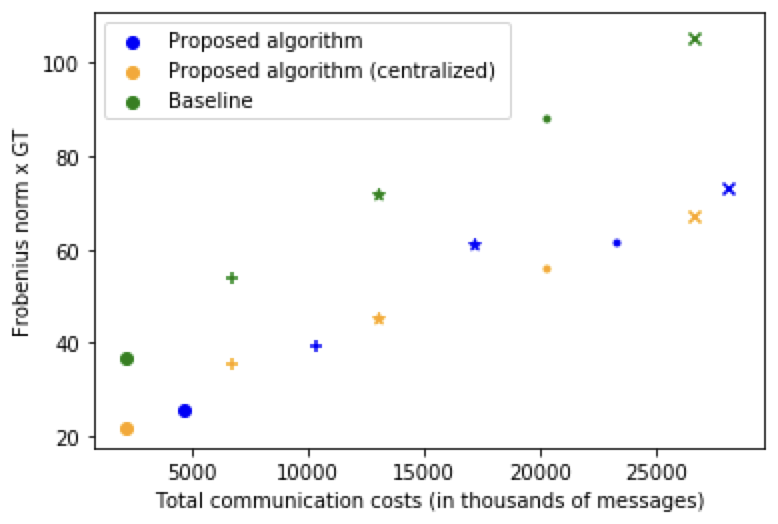}
	\caption{Frobenius norm  (of the difference between learned and ground truth weight matrices) versus total communication costs for the three methods in dense settings. Big circle: $N =150$, + symbol: $N=350$, star: $N=550$, small circle: $N=750$, X symbol: $N=950$.}
	\label{fig:FrobxCommDense}
\end{figure}

\begin{figure}[htb]
	\centering
	\includegraphics[width=\linewidth]{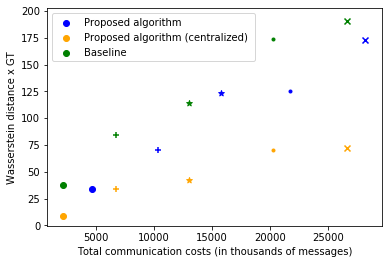}
	\caption{Wasserstein norm (between learned and ground truth weight matrices) versus total communication costs for the three methods in dense settings. Big circle: $N =150$, + symbol: $N=350$, star: $N=550$, small circle: $N=750$, X symbol: $N=950$.}
	\label{fig:WarsxCommDense}
\end{figure}

\paragraph*{Dense settings} 
In the same conditions as before, we use a bigger communication radius ( radius= $ \frac{3}{\sqrt{N}}$), which brings an average of 25.01 neighbours per node. We call these the dense conditions. We do the same experiments for the Frobenius norm and Wasserstein norm. For both norms, the higher density on the communication graph increases the communication costs for the distributed version and decreases costs for both centralised versions. This happens because the distances to the central node are lower when the nodes have more connections (for the centralized case). For the distributed case, it means the nodes have more neighbours to communicate to, which increases communication costs. In this experiment, the distributed method performs similarly to centralized methods, both in accuracy and communication costs. We can see the results for the Frobenius norm in dense settings versus the communication costs in Fig. \ref{fig:FrobxCommDense}, and for the Wasserstein norm in  Fig. \ref{fig:WarsxCommDense}.

\begin{figure}[htb]
		\centering
	\includegraphics[width=\linewidth]{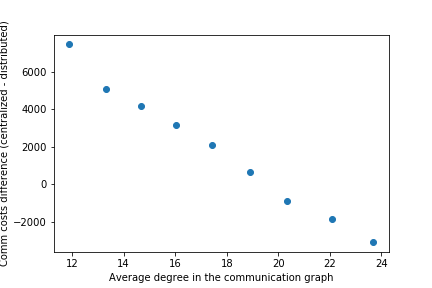}
	\caption{Communication costs difference between centralized and distributed methods against the average degree in the communication graph.   }
	\label{fig:deltaXsparsity}
\end{figure}

\paragraph*{Varying sparsity}
We now use 5000 signals, removal-rate equal to 0.7, we fix $N=500$ and we vary the communication radius from $\frac{2}{\sqrt{N}}$ to $\frac{3}{\sqrt{N}}$, consequently changing the communication graph sparsity. We plot the difference between the communication costs in the centralized and distributed versions against the average degree in the communication graph in Fig. \ref{fig:deltaXsparsity}. We observe that up to 20 neighbours per node, the communication cost in the distributed solution is lower. We use the normalized Frobenius norm, where we divide the weight matrices by their own Frobenius norms before calculating the norm of the difference to the ground truth. The observed differences between the normalized Frobenius norms of centralized and distributed versions, in Fig. \ref{fig:deltaXsparsity}, are very small, from left to right: 0.20, 0.14, 0.15, 0.13, 0.06, 0.02, 0.07, 0.06 and 0.04


\paragraph*{Changing the number of signals}

\begin{figure}[htb]
	\centering
	\includegraphics[width=\linewidth]{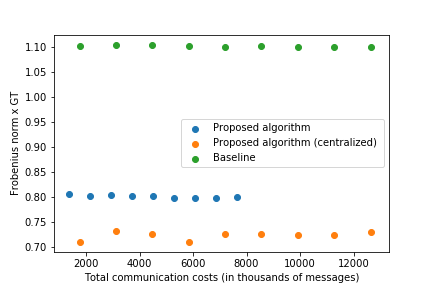}
	\caption{Normalized Frobenius norm (of the difference between learned and ground truth weight matrices)  versus total communication costs for the baseline method and both centralized and distributed proposed methods.}
	\label{fig:FrobxCommSignals}
\end{figure}

In the same settings as subsection \ref{sec:performance}, we fix N in 300, and vary the number of signals instead, from 1000 to 10000. In the results (Fig. \ref{fig:FrobxCommSignals}), we see that the normalized Frobenius does not change with the variation on the number of signals for the same 3 algorithms considered. For the baseline, the normalized Frobenius norm equals around 1.10, the centralized version of the proposed algorithm equals around 0.73 and the distributed version, 0.81.  The number of signals directly impacts the communication costs for all methods, as expected. The cost grows linearly with this number. We also note that the distributed version scales better with the growth of the number of signals.


\vspace{-1em}

\section{Conclusion}

We have proposed the first distributed graph learning framework that carefully considers communication costs. We model the problem with an objective function that has a low number of parameters and is simple to optimize. We solve it in two steps, an initialisation step and an optimisation step. In the latter, we use one independent optimiser per node, and we incorporate constraints with a mix of projection methods and a regularisation term. Our experiments show that the communication cost is lowered in the distributed algorithm without compromising the accuracy. Our solution scales better in terms of communication costs with the increase of the network size and the number of signals, compared to centralised solutions. We also show that sparse networks benefit more from a distributed solution than dense ones. Moreover, distributed solutions have other benefits by design, such as allowing the information to stay local, bypassing scheduling and transmission bottleneck issues and improving privacy and robustness against central node failure. 

\bibliographystyle{IEEEbib}
\bibliography{refs}

\end{document}